\documentclass[twocolumn,aps,prl,showpacs]{revtex4}
\renewcommand{\epsilon}{\varepsilon}
\usepackage{amsmath,epsfig,amsfonts,epsfig,graphicx,rotating
}

\begin{document}
\title{
Radiationless Travelling Waves In Saturable Nonlinear Schr{\"o}dinger Lattices}
\author{T.R.O. Melvin$^1$, A.R. Champneys$^1$, P.G. Kevrekidis$^2$ and J. Cuevas$^3$}
\affiliation{
$^1$ Department of Engineering Mathematics, University of Bristol,
BS8 1TR, UK\\
$^2$ Department of Mathematics and Statistics, University of
Massachusetts, Amherst, MA 01003-4515, USA \\
$^{3}$  Grupo de F{\'i}sica No Lineal, Departamento de
F{\'i}sica Aplicada I, Escuela Universitaria Polit{\'{e}}cnica,
Virgen de {\'{A}}frica, 7, 41011
Sevilla, Spain}

\begin{abstract}
The longstanding problem of moving discrete solitary waves 
in nonlinear Schr{\"o}dinger lattices is revisited.
The context is photorefractive crystal lattices with
saturable nonlinearity whose grand-canonical
energy barrier vanishes for isolated coupling strength values.
{\em Genuinely localised travelling waves}
are computed as a function of the system parameters {\it for the
first time}. The relevant
solutions exist only for finite velocities. 

\end{abstract}

\maketitle

Recently, the topic of discrete solitons (intrinsic localised modes)
in photorefractive materials has received significant attention, see
\cite{review05} for a review.  This interest was initialized by the
experimental realization of two-dimensional periodic 
lattices in photorefractive crystals 
\cite{ESCFS02} in which solitons were observed
\cite{FCSEC03,MECC04}. Further work has revealed a wealth of
additional coherent structures such as dipoles, quadrupoles, soliton
trains, vector, necklace and ring solitons, see
e.g.~\cite{YMKMMFC05,WCK06,NAOKMMC04,FBCMSHC04}.  
Since photorefractive materials
feature the so-called saturable nonlinearity,
these results for periodic lattices have spawned a parallel interest
in genuinely {\it discrete} saturable nonlinear lattices
\cite{kip,kip2,cuevas}.  One particularly intriguing result of theses
studies is that the stability properties of the localized modes are
substantially different from their regular discrete nonlinear
Schr{\"o}dinger (DNLS) analogs. In DNLS it is well-known
\cite{dnls,dnls2,non} that site-centered localised modes are always
stable, while intersite-centered modes are unstable (and are only
stabilized in the continuum limit where these two branches degenerate
into the well-known continuum sech-soliton of the integrable cubic
NLS). For the photorefractive nonlinearity (i.e., the so-called
Vinetskii-Kukhtarev model originally proposed in \cite{VK} and
revisited in \cite{kip,kip2,cuevas}), depending on the coupling
strength, the inter-site centered modes may have lower energy than
their onsite counterparts. Hence, one should expect that the ensuing
sign reversal of the so-called Peierls-Nabarro (PN) energy barrier
$\Delta E=E_{IS}-E_{OS}$ (where the subscripts denote intersite and
onsite respectively) should cause an exchange between the linear
stability properties of the two modes.

A related, even more fundamental, question in nonlinear lattice models
of DNLS type is whether exponentially localized self-supporting
excitations that move with a constant wave speed can exist, so called
{\em moving discrete solitons}. For continuum models this question is
in some sense trivial since the equations posed in a moving frame
remain of the same fundamental type. Yet for lattices, passing to a
moving frame leads to so-called advance-delay equations that are
notoriously hard to analyse. One recent (negative) result in this
direction \cite{floria2} for the so-called Salerno model, shows that,
starting from the integrable Ablowitz-Ladik equation, mobile discrete
solitary waves acquire non-vanishing tails as soon as parameters deviate
from the integrable limit. Hence exponentially localised fundamental
(single humped) moving discrete solitons cannot be constructed.
While these results settle a long-standing controversy, see
e.g.~\cite{stuff1,stuff2,stuff3}, they are also somewhat
unsatisfactory since they do not give conditions under which moving
discrete solitons might exist for generic, non-integrable lattices.

In the arguably simpler problem of kinks in 
Frenkel-Kontorova lattices
(and, more generally, in
discrete Klein-Gordon models),
genuinely travelling fundamental (`charge one') topological solitons
are known {\em not} to occur unless there is a ``competing''
nonlinearity that causes vanishings of the PN barrier;
see e.g.~\cite{savin,aigner}. Drawing an analogy from this,
for (the more complicated, complex field) 
DNLS models with pure cubic nonlinearity,
the PN barrier never disappears, so one should not expect genuinely
localised moving solitons (in some sense anticipating the
above-mentioned negative result).  However the present saturable model
has a nonlinear term in its denominator, which, under Taylor
expansion, is effectively cubic for low intensities but exhibits
saturation for large intensities. This effective ``competition'' between
nonlinear behaviors leads to the possibility of the vanishing of the PN
barrier, and hence suggests this model as  a good candidate for
finding moving discrete solitons. Such a search is the theme of
this Letter. We note that such a connection was also partially
discussed in \cite{maluckov}, based on the properties of the 
{\it stationary} localized modes.

We start by examining the existence of stationary localized modes as
the strength of inter-site coupling is varied, initializing our
computations from the anti-continuum limit, where explicit solutions
are available.  A linear stability analysis of the solutions is
carried out, and we find that the regime of stability-instability
alternation does not coincide with the vanishings of the PN barrier
but rather with the vanishings of the grand-canonical free-energy of
the system $G=E-\Lambda P$, where $\Lambda$ is the frequency (chemical
potential) of the solution and $P$ is its $l^2$ norm.  After
identifying this sequence of {\em transparent points}, we proceed to
obtain, for the first time in discrete nonlinear Schr{\"o}dinger type
models, genuinely localized, single-humped, uniformly traveling
solutions in their vicinity. This is done by examining travelling
solutions as embedded solitons \cite{yang}, such that while resonant with the
linear spectrum in the travelling frame, the tail amplitude can vanish
for isolated, definitely non-zero values of the speed, for a given set
of system parameters. This is reminiscent of recent work in
the Klein-Gordon context \cite{OBP}, where the so-called
Stokes constant is shown to vanish for isolated values of
the speed, indicating the presence of travelling localised solutions.



The model of interest, as discussed above, will be a
discrete lattice with photorefractive nonlinearity, namely,
\begin{eqnarray}
i \dot{u}_n=-\epsilon \Delta_2 u_n + \frac{\beta}{1+ |u_n|^2} u_n.
\label{peq1}
\end{eqnarray}
Here, $\epsilon$ denotes the coupling strength between adjacent sites
and $\beta$ is the coefficient of the nonlinearity (proportional
to the voltage in photorefractive crystals) and can be scaled out
for our purposes; $\Delta_2 u_n=u_{n+1}+u_{n-1}-2u_n$ denotes the discrete
Laplacian.
We will therefore set $\beta=1$ in what follows.
We seek stationary solutions of the form
$u_n=e^{-i \Lambda t} v_n$,
starting from the so-called
anti-continuum (AC) limit $\epsilon=0$. In that limit the principal
solutions of interest are the onsite solution with
$u_{n_0}=\pm \sqrt{(1/\Lambda)-1}$,
(all other sites bearing vanishing amplitudes)
and the intersite-centered solution with
$u_{n_0}=u_{n_0+1}=\pm \sqrt{(1/\Lambda)-1}$
and zero amplitudes elsewhere.
We initialize these exact profiles from the
AC limit and subsequently use continuation in $\epsilon$ to
solve the stationary equations for the wave profile
via Newton's method for finite $\epsilon$.
We also examine the linear stability of the obtained modes using
linearization around a solution profile $v_n^0$ of the form:
$u_n=e^{-i \Lambda t} \left[v_n^0 + \delta \left(a_n e^{\lambda t} +
b_n e^{\lambda^{\star} t} \right) \right]$,
and resolving the ensuing matrix eigenvalue problem ensuing to
O$(\delta)$ for the eigenvalue $\lambda$ and the eigenvector
pair $(a_n,b_n^{\star})$ (where $^\star$ stands for complex conjugation).
In what follows, we will also use the (Hamiltonian) energy of solutions,
\begin{eqnarray}
E=\sum_n \left[ \epsilon |u_{n+1}-u_n|^2 + \log(1+|u_n|^2) \right],
\label{peq5}
\end{eqnarray}
and the $l^2$-norm, $P=\sum_n |u_n|^2$.


\begin{figure}[tbp]
\begin{center}
\epsfxsize=4.25cm 
\epsffile{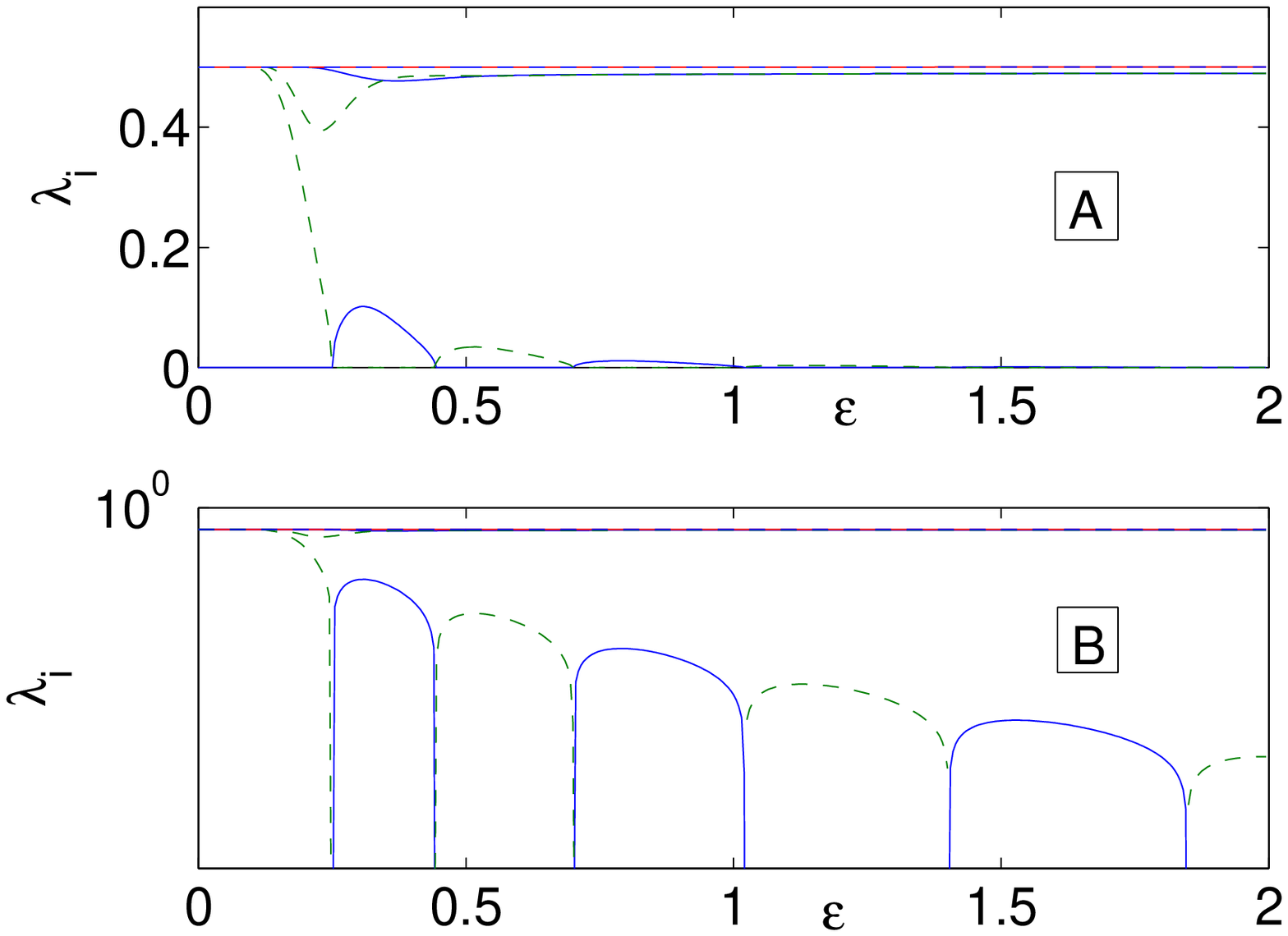}
\epsfxsize=4.25cm 
\epsffile{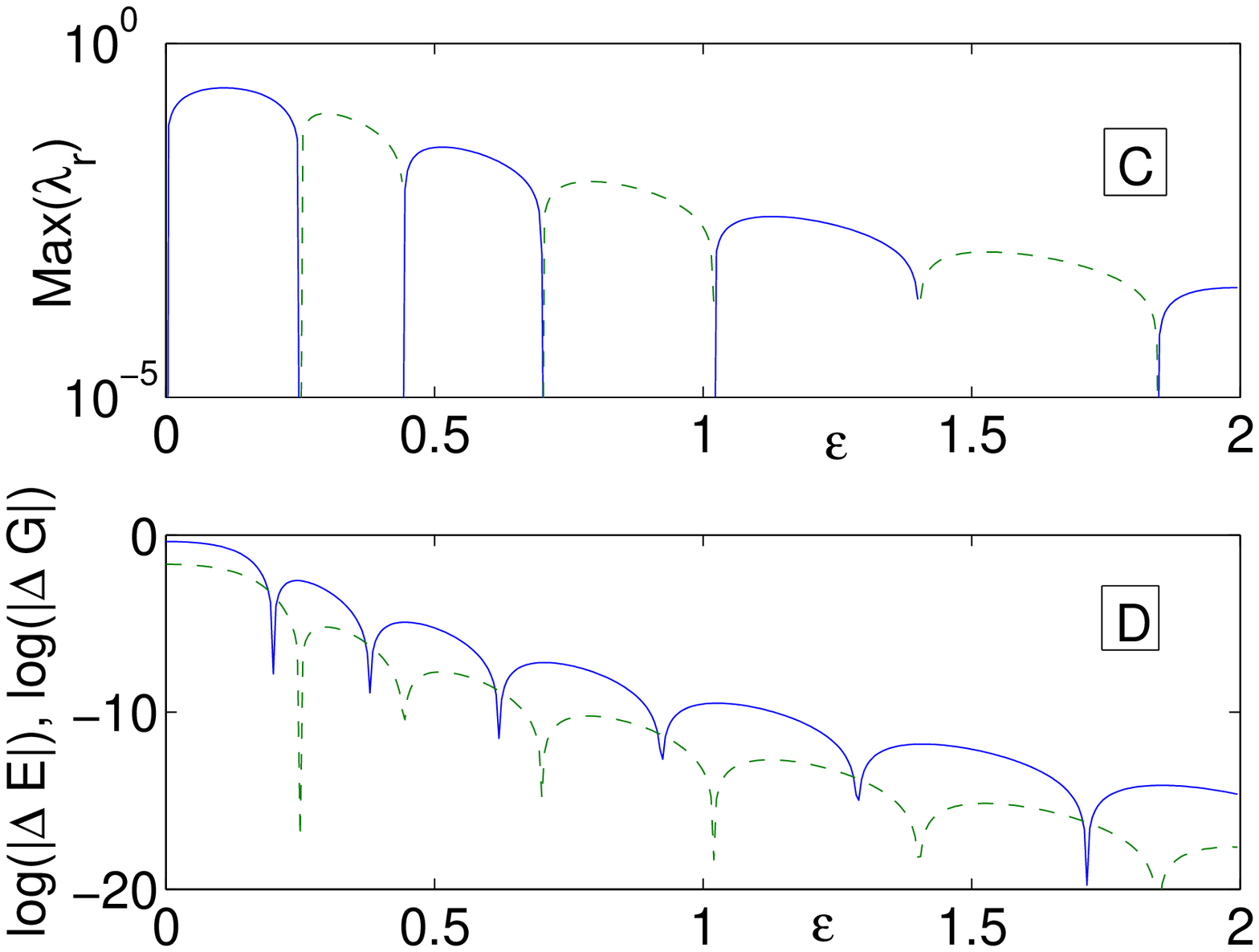}
\caption{(Color Online) Panels A and B 
show the key imaginary eigenvalues for the onsite
(dashed line) and the intersite (solid line) mode as a function of
$\epsilon$ (A on a linear scale, B on an exponential scale).
The band edge of the continuous spectrum is
at $\Lambda=0.5$. From there bifurcates an eigenmode for
$\epsilon>0.1$ (for the onsite case) which arrives at the origin
of the spectral plane for $\epsilon=0.25$ and becomes real. On the
contrary for the intersite solution a previously real eigenmode
exits as imaginary for $\epsilon>0.25$. These modes alternate again
for $\epsilon=0.445$, $\epsilon=0.7$ etc. The real parts of the
corresponding eigenvalues are shown in the panel C.
Panel D shows $\log(|\Delta E|)$
between onsite and intersite modes, and by dashed line the 
quantity $\log(|\Delta G|)$, where $G=E-\Lambda P$ (see
text for details).}
\label{pfig1}
\end{center}
\end{figure}

Our stationary calculations as a function of the inter-site coupling
(rather than of the frequency as in \cite{kip}) are shown in
Fig.~\ref{pfig1}.  Panels A and B show the key
imaginary eigenvalues of the problem, while panel C
shows the maximal real eigenvalues. 
A solid line is used in the figures to represent
the inter-site centered (IS) mode, and a dashed line the
onsite-centered (OS) mode.  Observe that for small values of
$\epsilon$ the OS mode is stable while the intersite mode is
unstable. Consider the OS mode. At $\epsilon \approx 0.1$, an
eigenvalue emerges from the band edge of the continuous spectrum at
$\lambda=i \Lambda$, and approaches the origin of the spectral plane.
This feature also occurs in the pure cubic DNLS but the eigenvalue
only arrives at the origin of the spectral plane as $\epsilon
\rightarrow \infty$ (the continuum limit).  Here the eigenvalue
arrives at the origin for $\epsilon \approx 0.25$, becomes real for
$0.25<\epsilon<0.445$, then becomes imaginary again for
$0.445<\epsilon<0.7$, then real again, etc.  Furthermore, this
alternation of stability occurs hand-in-hand with the alternation of
stability of the IS mode, which starts out unstable, becomes stable,
then unstable again, with the transitions occurring at precisely the
same $\epsilon$-values.  Note from the figure that stability
alternation occurs, {\it not} at the points of vanishing of the PN
energy barrier; but when the {\it grand-canonical energy barrier} of
the model $\Delta G=\Delta (E-\Lambda P)$ between the IS and OS modes 
vanishes. This
provides an alternative viewpoint to the results of \cite{kip},
which is also rather natural from the point of view of statistical
mechanics for a system with a conserved ``particle number'' ($l^2$ norm); 
see e.g. \cite{leb}. 
At these very points, the relevant eigenvalue (pertaining to translation)
crosses zero, instantaneously restoring an effective translational
invariance in the model. Thus these are {\em transparent points} where
there is a one-dimensional family of localised solutions, obtained
from each other by translation, of which the OS and IS modes are just
two examples.

\begin{figure}[tbp]
\begin{center}
\epsfxsize=8cm 
\epsffile{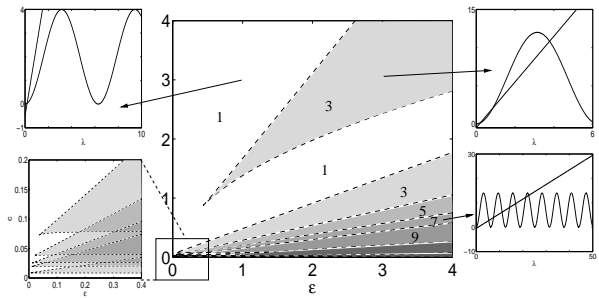}
\caption{Intersections of the single and double root conditions
(see text) for  $\Lambda = 0.5$ and varying $c$ and $\epsilon$. The
shaded area shows where there is more than one branch of linear waves.
Values indicate the number of roots of (\ref{specPdnls2}). Subplots show
left and right hand sides of (\ref{specPdnls2}) for ($\epsilon,c$)  = (1,3),
one root (i), (3,3), three roots (ii) and (3.5,0.6),
seven roots (iii). (iv) displays the overlapping of the bands for
small $\epsilon$ and $c$, only the first six bands have been shown.}
\label{pfig2}
\end{center}
\end{figure}

Intuitively, one might expect that the transparent points represent
parameter values at which genuinely travelling localised solutions
of the photorefractive lattice might bifurcate with zero wave speed.
Hence we shall look for solutions to Eq.~(\ref{peq1}) near such
points using a travelling wave substitution $u_{n}(t) =
\psi(n-ct)e^{i(k n - \Lambda t)}$.  This leads to the advance-delay
equation in the travelling co-ordinate $z=n-ct$
\begin{eqnarray}
\label{twPdnls}
-ic\psi'(z) &=& (2\epsilon-\Lambda)\psi(z) - \epsilon(\psi(z+1)e^{ik} + \psi(z-1)e^{-ik})
\nonumber
\\
&+& \frac{1}{1+\left|\psi(z)\right|^{2}}\psi(z),
\end{eqnarray}
where $'$ denotes differentiation with respect to $z$.
Equation (\ref{twPdnls}) is rotationally invariant, therefore the
transformation
$\psi(z) = \tilde{\psi}(z)e^{-ikz}$
can be used to obtain the same equation as above for $\tilde{\psi}$ and
$\tilde{\Lambda}=kc+\Lambda$. Therefore, $k = 0$ will be used hereafter
in  (\ref{twPdnls}) without loss of generality.
The linear spectrum is obtained by substitution of
$\psi(z) = e^{i\lambda z}$ into the linear part of Eq.~(\ref{twPdnls})
\begin{eqnarray}\label{specPdnls2}
c\lambda + \Lambda - 1 = 4\epsilon\sin^{2}(\frac{\lambda}{2})
\end{eqnarray}
It is interesting to note that the asymptotic behaviour of the two
sides of Eq.~(\ref{specPdnls2}) for small $\lambda$
is such that a resonance with the linear spectral bands is unavoidable.
However, in the hope of obtaining
a localized solution such resonances need to be minimized and hence
we need to investigate parameter regions where Eq.~(\ref{specPdnls2})
has only one root, $\lambda > 0$. This can be efficiently done \cite{aigner}
by considering the conditions for double roots, namely
$c = 2\epsilon\sin(\lambda)$, which allow us to distinguish among regions
with different root multiplicity. The results are plotted in
Fig.~\ref{pfig2}, from which it can be seen that for each $\epsilon>0$,
the number of spectral bands increases to infinity as $c\to 0$.
This result serves to counter the
naive expectation that near the transparency points, travelling solutions
of small speed may exist.

The best hope for finding genuinely localised solutions is then within
parameter regions of Fig.~\ref{pfig2} in which there is a single
resonance with linear waves. Such solutions would represent
embedded soliton structures \cite{yang,aigner} for which
the radiation mode component exactly vanishes in the tail of the
solitary wave. At best this could happen in a codimension-one set
in parameter space, that is along discrete curves lying within
the single resonance bands of Fig.~\ref{pfig2}.
Solutions of this type are sought
using a pseudo-spectral method, see \cite{stuff2,savin}. To this end, we set
\begin{eqnarray}\label{PSsub}
\psi(z) = \sum^{N}_{j=1} a_{j}\cos(\omega_{j}z) + ib_{j}\sin(\omega_{j}z)
\end{eqnarray}
with $\omega_{j} = \frac{\pi j}{L}$, to transform (\ref{twPdnls})
into a system of algebraic equations in the long finite interval
$[-L/2, L/2]$. $a_{j}, b_{j} \in {\mathbb R} $ are the coefficient
of the Fourier series, for which the ensuing algebraic equations are
solved at the collocation points $z_{m} = \frac{Lm}{2(N+1)}$. The
solution is obtained by means of the  Powell hybrid method
\cite{powell}, with an error tolerance of $10^{-13}$. From the
resulting $a_j, b_j$'s, $\psi(z)$ can be reconstructed and
numerically continued using AUTO \cite{doedel} with an error
tolerance of $10^{-10}$ to investigate the effect of varying
parameters on the solution shape and its tail amplitude. Generally,
the thus obtained solutions will be 
weakly nonlocal ones with
non-zero oscillatory tails. However, appending an additional tail
condition for the solution, yielding a signed measure $\Delta$ of
the amplitude of the tail, we can obtain truly localised solutions
as isolated zeros of this function $\Delta$. Given our choice of
phase in the substitution (\ref{PSsub}) that the real component of
$\psi$ is odd about $z=L/2$ (and vice versa for the imaginary
component), a good choice for such a tail amplitude measure is
$\Delta = \rm{Im}(\psi(\frac{L}{2}))$.

\begin{figure}[tbp]
\begin{center}
\epsfxsize=4cm 
 \epsffile{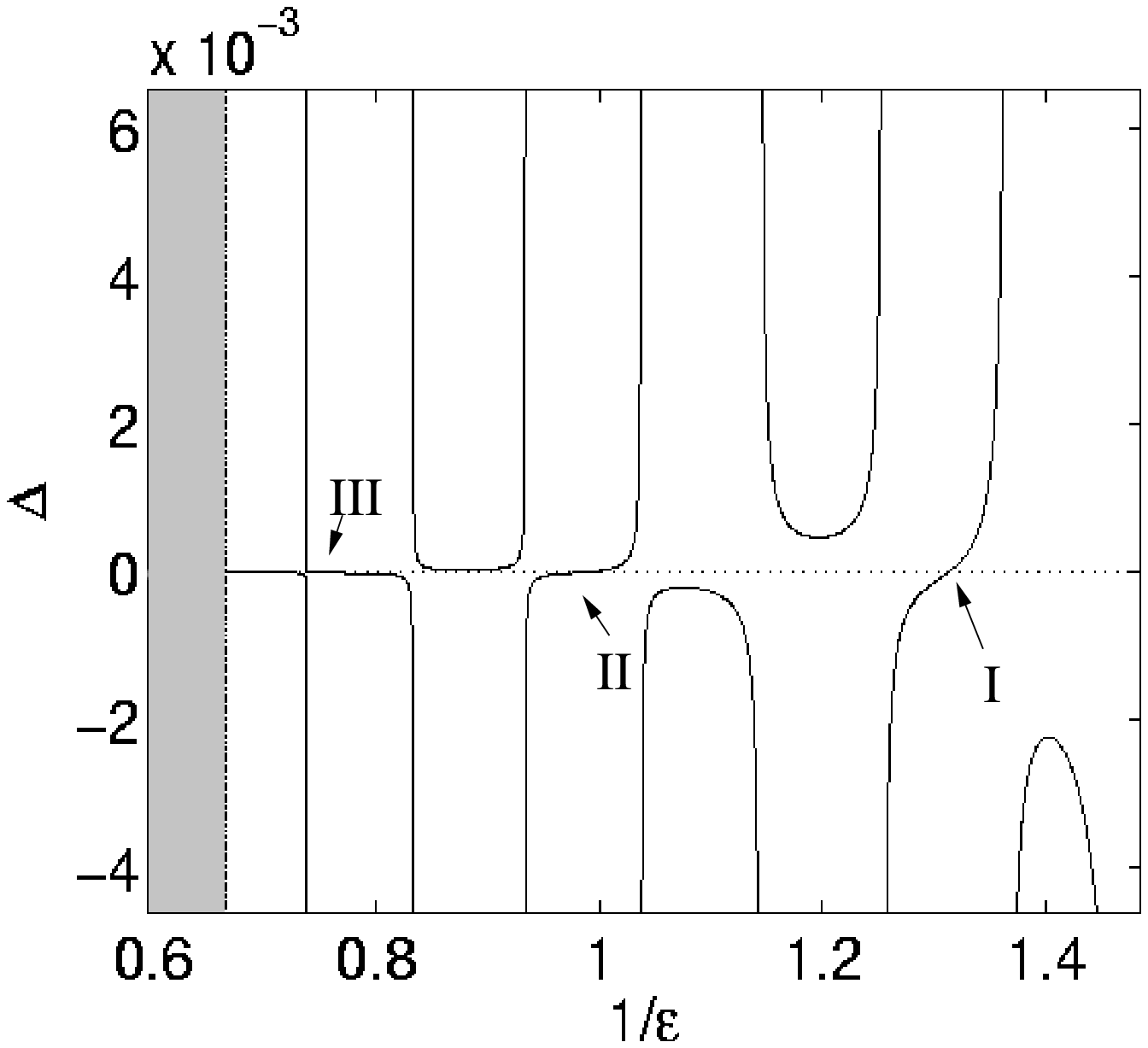}
\epsfxsize=4cm 
 \epsffile{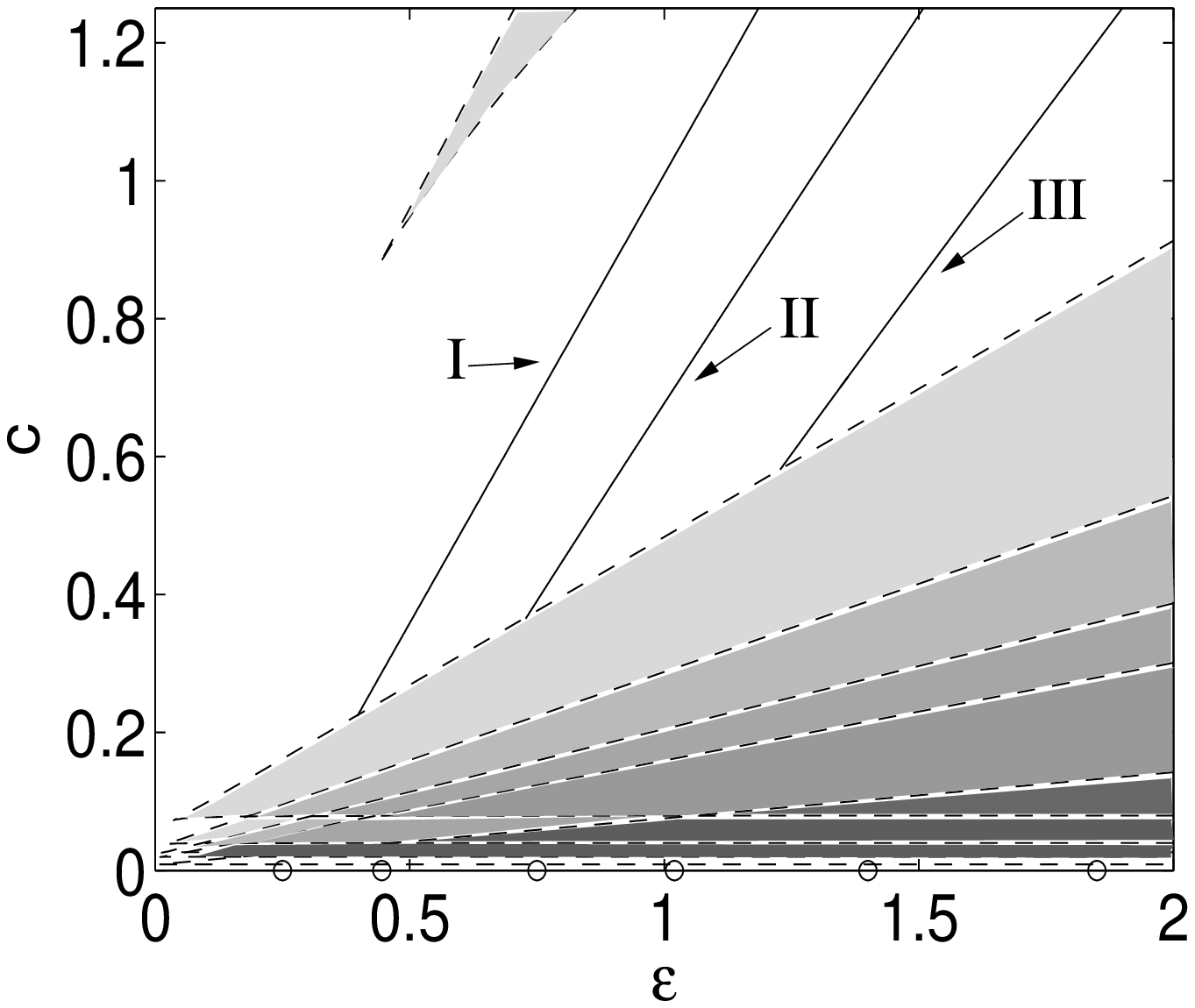}
\epsfxsize=4cm 
 \epsffile{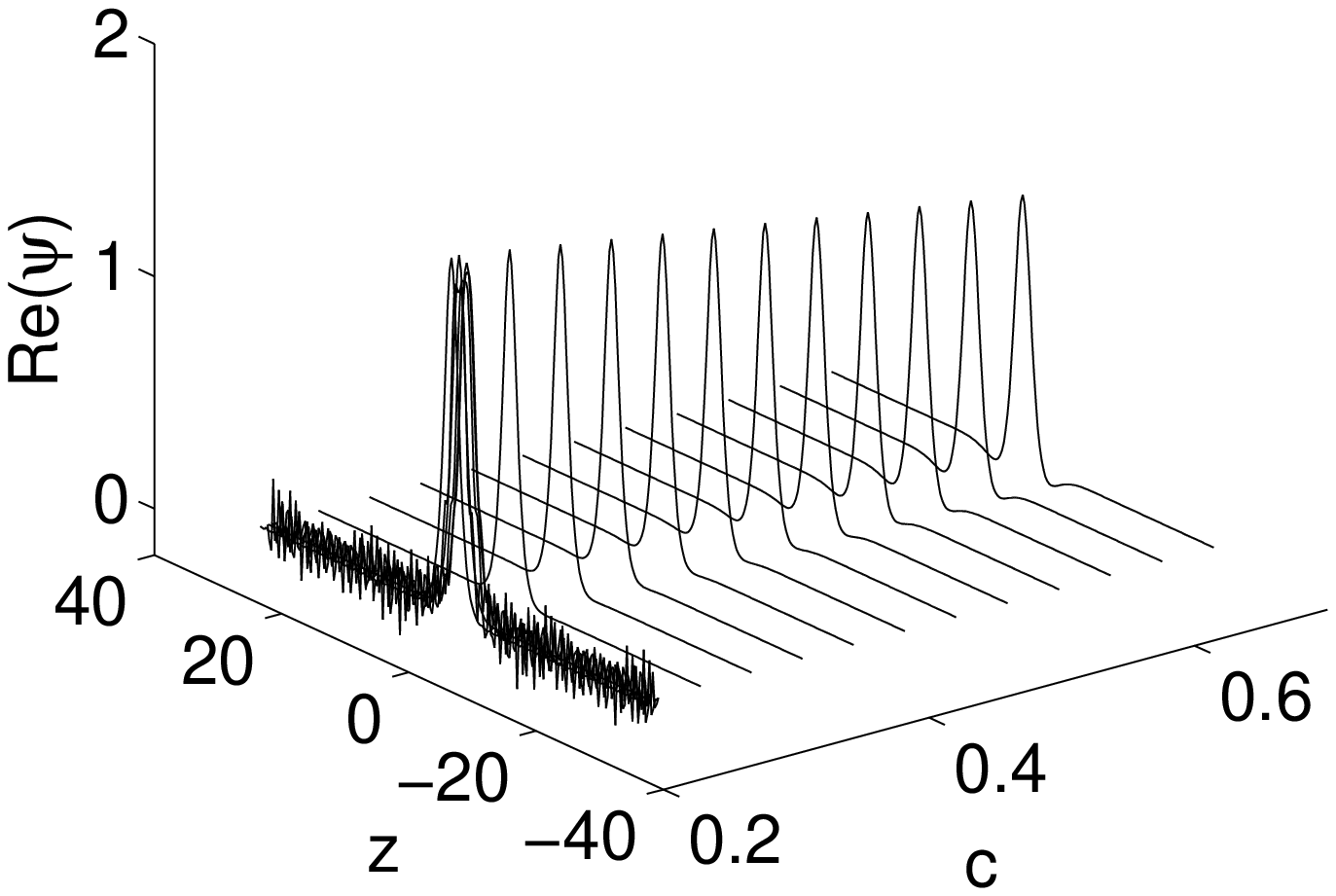}
\epsfxsize=4cm 
 \epsffile{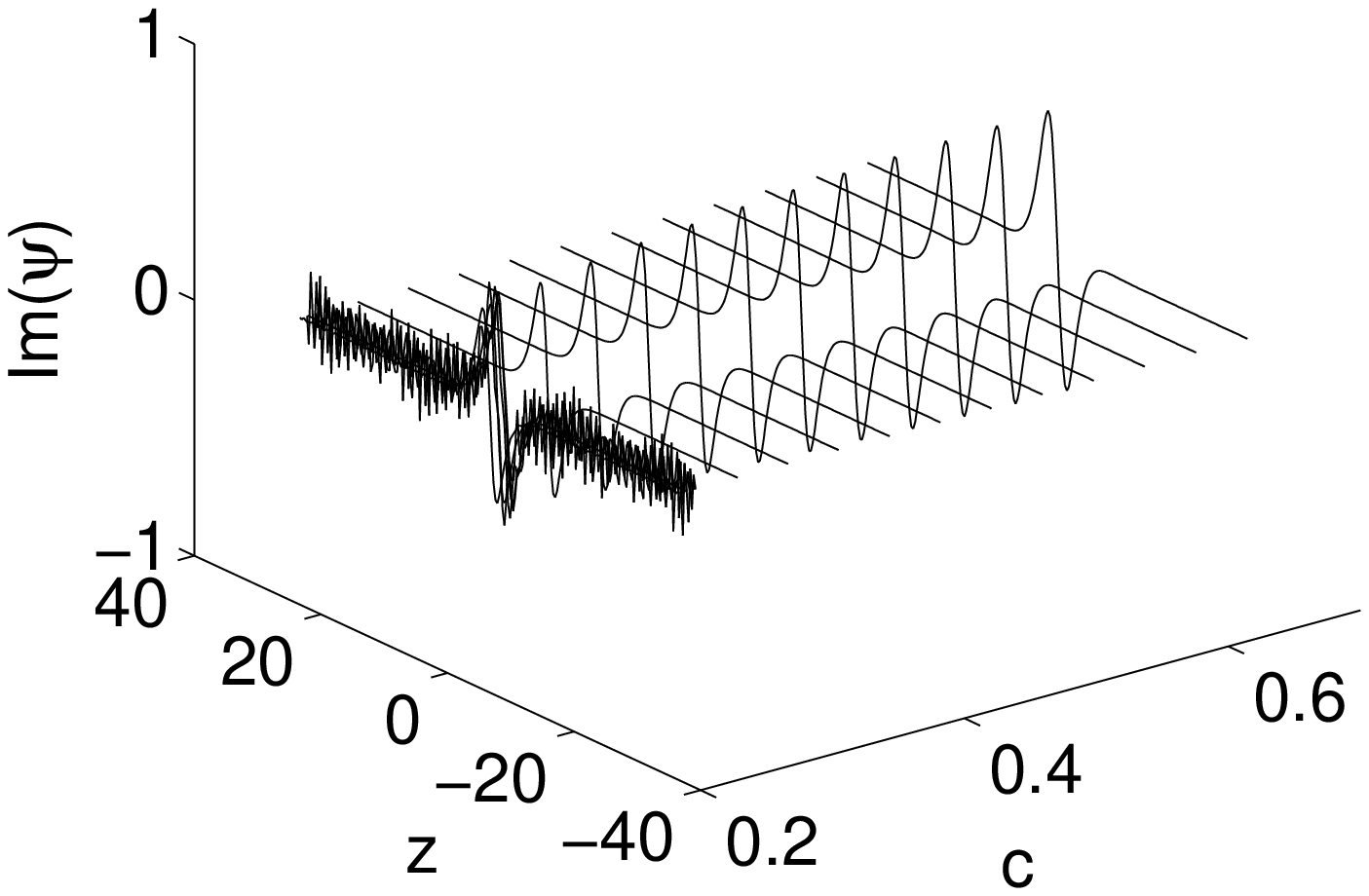}
\epsfxsize=4cm 
 \epsffile{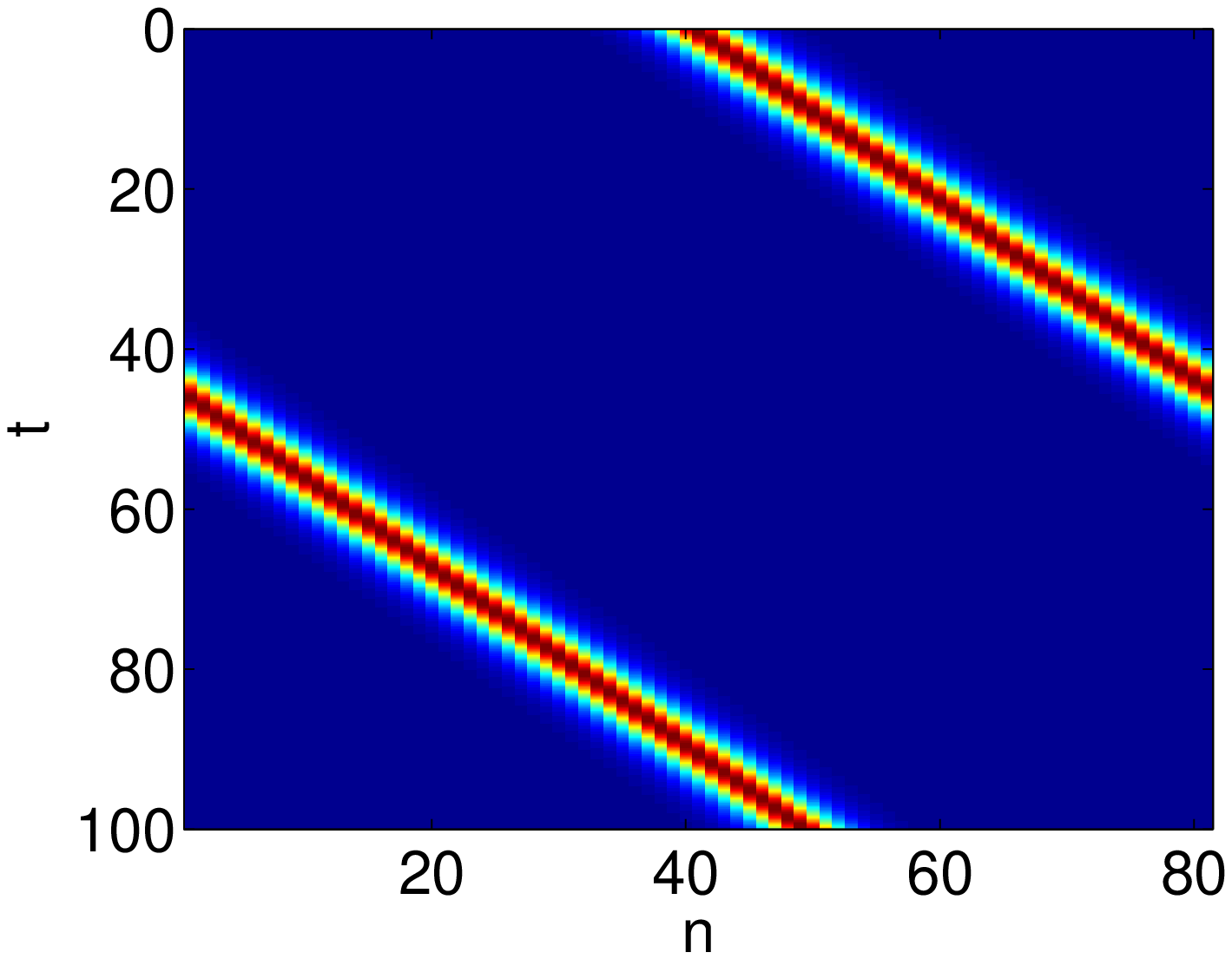}
\epsfxsize=4cm 
 \epsffile{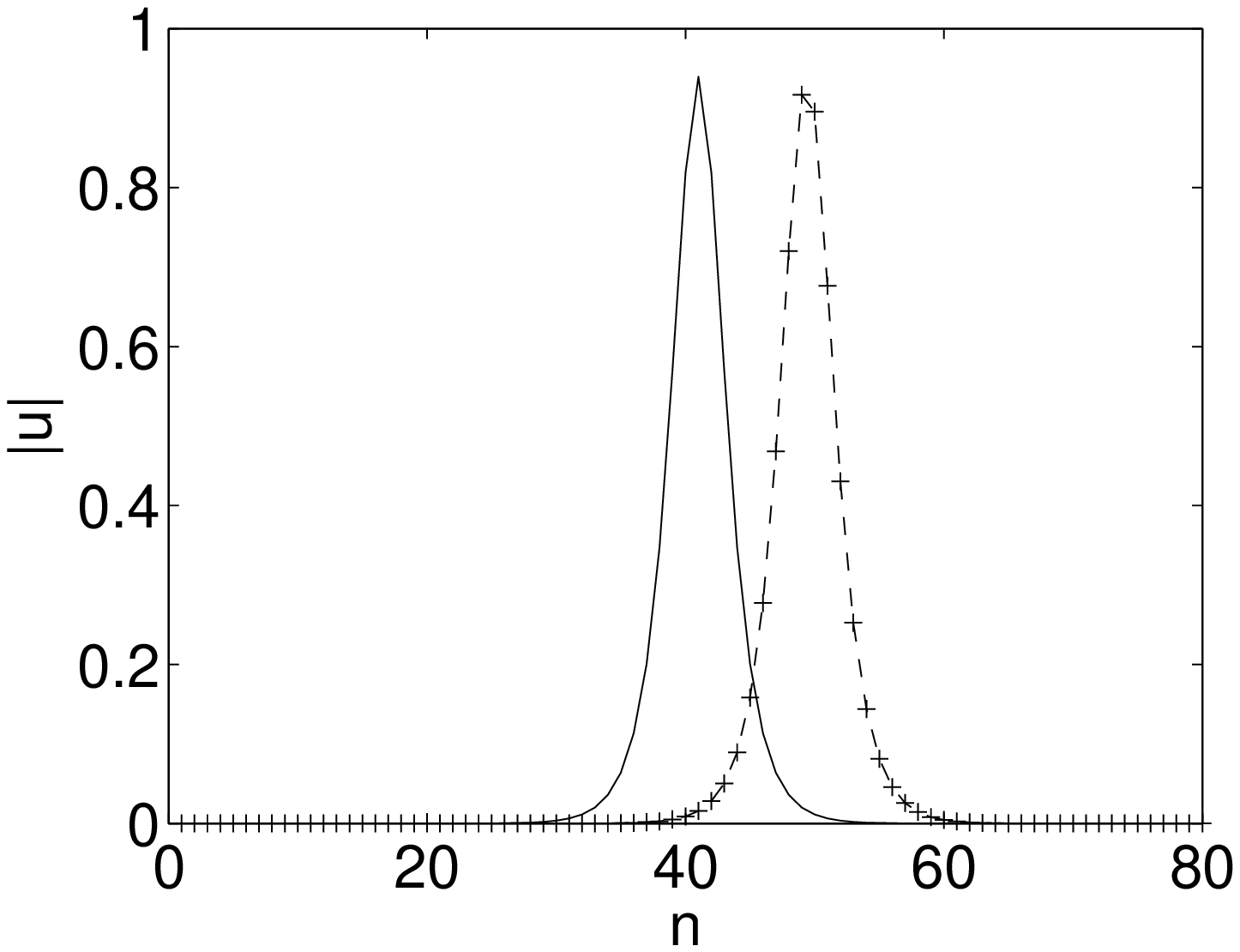}

\caption{(Top left) Continuation of weakly nonlocal solitary waves for various values of $1/\epsilon$ showing three zeros in $\Delta$ for $c = 0.7, \Lambda = 0.5, L = 60$. Zeros of $\Delta$ at $\epsilon \approx 0.76, 1.02, 1.36$. (Top right). Continuation of the 3 zeros of $\Delta$, varying $\epsilon$ and $c$ with $\Lambda = 0.5$. Circles on the $c = 0$ axis indicate the transparency points.
Middle 
panels show the continuation
of the (real and imaginary parts of the) solution of branch I for 
values of the speed. Examples of direct integration of the solution of
branch I with $\epsilon = 0.911396$, $c = 0.894153$ (bottom panels).
The space time contour plot of the solution modulus (bottom left panel),
and the modulus before (solid) and after (dashed)
100 time steps (bottom right panel) are shown.}
\label{pfig3}
\end{center}
\end{figure}

Continuation of weakly nonlocal solitary wave solutions as $\epsilon$ varies is
shown as a function of $1/\epsilon$ in Fig.~\ref{pfig3} (top left).
Note the topological distinction between branches the `S'-shaped
branches that contain true zeros of $\Delta$ and the `U' or
`n'-shaped branches that do not \cite{vandenbroek}. Once the
isolated points where the tail disappears are found, these can be
continued in any pair of desired parameters, e.g.~ $(\epsilon,c)$ by
appending the condition $\Delta=0$ to the numerical problem. Thus,
we can construct the parametric diagram of the existence of such
{\it genuinely travelling} (with constant prescribed speed), yet
{\em truly localized} (i.e., radiationless 
due to $\Delta=0$) solutions. A regular
sequence of branches so-obtained are found as $\epsilon$ increases,
the first three of which are shown in the top right panel  of
Fig.~\ref{pfig3}. Notice that each branch terminates (as it should)
at the edges of the multi-spectral bands. Nevertheless, at the
low-$c$ end the branches are close to straight lines which if
continued to zero wave speed would hit the transparent points. Thus,
while there is no actual bifurcation as such of intrinsic localised modes
from these transparent points, their presence has a strong
influence.

The 
middle two panels of Fig.~\ref{pfig3} show solution profiles on
the first (lowest $\epsilon$) branch of localised solutions.  Note
how the profile delocalizes when the edge of the
multi-resonance band is reached. Any attempt to continue zero-tail
solutions into the band resulted in lack of convergence 
for large
enough $L$ and $N$.
The lower two panels show the result
of direct integration of a representative solution on 
branch I, clearly illustrating their genuinely travelling
nature, and indicating that they are stable against the perturbation
introduced by numerical discretization.

In conclusion, to our knowledge, this is the {\it first time}
genuinely localised travelling waves 
have been found in a discrete lattice of nonlinear
Schr{\"o}dinger type. While it is true that our construction has
relied on numerical continuation, we have demonstrated the robustness
of the solutions we have obtained, and have provided a new rationale
as to why they should exist. This rationale is based on the existence
of transparent points for stationary localised modes when a
generalised Peierls-Nabarro barrier vanishes. However, we have also
shown that this is not enough; one also needs the wave speed to be
sufficiently large to overcome the multi-spectral bands. Within each
band, exponentially localised solutions occur as embedded solitons
which correspond to {\em regular zeros} of an appropriately defined
measure of the tail  amplitude.

We anticipate that the principal features of the model problem studied
(competing
nonlinear terms for the existence of transparent points and sufficiently
high wave speeds to overcome the multi-spectral bands) should be generic
features a variety of similar Hamiltonian lattices. Indeed, we have found
 similar results for a pure cubic-quintic nonlinearity. More
realistic models of photorefractive crystals should be expected to behave
similarly, giving the possibility of experimental observation of
the waves we have constructed. There, the
robust, radiationless nature of the new moving pulses
may be particularly appealing.
Another relevant question is the extension of the
present ideas to two-dimensional settings \cite{vicencio}, where
deciding which directions of wave propagation within the lattice
can lead to radiationless pulses will also be of interest.


\begin{thebibliography}{99}

\bibitem{review05} J.W. Fleischer,
Opt. Express {\bf 13}, 1780 (2005).


\bibitem{ESCFS02} N.K. Efremidis {\it et al.},
Phys. Rev. E {\bf 66} (2002)
046602.


\bibitem{FCSEC03} J.W. Fleischer {\it et al.},
Phys. Rev. Lett. {\bf 90}
(2003) 023902.

\bibitem{MECC04} H. Martin {\it et al.},
Phys. Rev. Lett. {\bf 92} (2004) 123902.




\bibitem{YMKMMFC05} J. Yang {\it et al.},
Phys. Rev. Lett. {\bf 94}, 113902 (2005).

\bibitem{WCK06} X. Wang, Z. Chen and P.G. Kevrekidis,
Phys. Rev. Lett. {\bf 96}, 083904 (2006).

\bibitem{NAOKMMC04} D.N. Neshev {\it et al.},
Phys. Rev. Lett. {\bf 92} (2004) 123903.

\bibitem{FBCMSHC04} J.W. Fleischer {\it et al.},
Phys.
Rev. Lett. {\bf 92} (2004) 123904.




\bibitem{kip} L. Hadziewski {\it et al.}, Phys. Rev. Lett.
{\bf 90}. 033901 (2004).

\bibitem{kip2} M. Step{\'i}c {\it et al.}, Phys. Rev. E {\bf 69}, 066618
(2004).

\bibitem{cuevas} J. Cuevas and J.C. Eilbeck, nlin.PS/0501050.

\bibitem{dnls} P.G. Kevrekidis, K.{\O}. Rasmussen and A.R. Bishop,
Int. J. Mod. Phys. B {\bf 15}, 2833 (2001);
\bibitem{dnls2} D. N.\ Christodoulides,
F.\ Lederer, and Y.\ Silberberg, Nature \textbf{424}, 817 (2003);


\bibitem{non} T. Kapitula and P. Kevrekidis,
Nonlinearity {\bf 14}, 533 (2001).

\bibitem{VK} V.L. Vinetskii and N.V. Kukhtarev,
Fizika Tverdogo Tela {\bf 16}, 3714 (1974).




\bibitem{floria2} J. G{\'o}mez-Garde\~{n}es, L.M. Flor{\'i}a,
M. Peyrard and A.R. Bishop, Chaos {\bf 14}, 1130 (2004).


\bibitem{stuff1} M. Remoissenet and M. Peyrard (eds.)
{\it Nonlinear Coherent Structures in Physics and Biology} Springer-Verlag
Berlin, (1991);
\bibitem{stuff2} D.B. Duncan {\it et al.}, Physica D {\bf 68}, 1 (1993);
%
\bibitem{stuff3}
M.J. Ablowitz, Z.H. Musslimani and G. Biondini,
Phys. Rev. E {\bf 65}, 026602 (2002).

\bibitem{savin} A.V. Savin, Y. Zolotaryuk and J.C. Eilbeck,
Phys. D. {\bf 138}, 267 (2000).

\bibitem{aigner} A Aigner, A.R. Champneys and V.M. Rothos
Phys. D {\bf 186}, 148 (2003).

\bibitem{maluckov} A. Maluckov, Lj. Hadzievski and M. Stepic
Phys. D {\bf 216}, 95 (2006).

\bibitem{yang} J. Yang, B.A. Malomed and D.J. Kaup,
Phys. Rev. Lett. {\bf 83}, 1958 (1999).

\bibitem{OBP} O.F. Oxtoby, D.E. Pelinovsky and I.V. Barashenkov,
Nonlinearity {\bf 19}, 217 (2006).

\bibitem{leb} J.L. Lebowitz, H.A. Rose and E.R. Speer,
J. Stat. Phys. {\bf 50}, 657 (1988); K.{\"O}. Rasmussen
{\it et al.}, Phys. Rev. Lett. {\bf 84}, 3740 (2000).

\bibitem{powell}  M.J.D. Powell.
\textit{Numerical Methods for Nonlinear Algebraic Equations.}
(Gordon and Breach, 1970).

\bibitem{doedel}  E.J. Doedel {\it et al.},
AUTO97 continuation and bifurcation software for ordinary
differential equations, available via:
\textit{ftp://ftp.es.concordia.ca/directory/doedel/auto}

\bibitem{vandenbroek} A.R. Champneys, J.-M. Vanden-Broeck and G.J. Lord,
{\em J. Fluid Mech}, {\bf 454}, 403 (2002).

\bibitem{vicencio} R. Vicencio, M. Johansson, Phys. Rev. E 
{\bf 73}, 046602 (2006). 

\end{thebibliography}
\end{document}